\documentclass[letterpaper,12pt]{article}

\usepackage[dvips]{graphicx}
\usepackage[pdftex]{color}
\usepackage{ifthen}
\usepackage{enumerate}
\usepackage{cite}
\usepackage{amssymb}
\usepackage[mathscr]{eucal}
\usepackage[errorshow]{tracefnt}
\usepackage{amsmath}
\usepackage{amssymb}
\usepackage{amsthm}
\usepackage{eucal}
\usepackage{eufrak}

\newcommand{\bfU}{\ensuremath{\boldsymbol{U}}}
\newcommand{\dlb}{\ensuremath{[\![}}
\newcommand{\drb}{\ensuremath{]\!]}}

\newcommand{\vG}{\ensuremath{\mathcal{G}}}

\newcommand{\beq}{\begin{equation}}
\newcommand{\beqn}{\begin{equation*}}
\newcommand{\eeq}{\end{equation}}
\newcommand{\eeqn}{\end{equation*}}
\newcommand{\beqa}{\begin{eqnarray}}
\newcommand{\beqan}{\begin{eqnarray*}}
\newcommand{\eeqa}{\end{eqnarray}}
\newcommand{\eeqan}{\end{eqnarray*}}
\newcommand{\bdm}{\begin{displaymath}}
\newcommand{\edm}{\end{displaymath}}

\newcommand{\la}{\langle}
\newcommand{\ra}{\rangle}

\newcommand{\ba}{\begin{array}}
\newcommand{\ea}{\end{array}}

\newcommand\ffam{\sffamily}
\newcommand\fser{\bfseries}

\newcommand\nn{\nonumber}

\newcommand\benu{\begin{enumerate}}
\newcommand\eenu{\end{enumerate}}
\newcommand\bit{\begin{itemize}}
\newcommand\eit{\end{itemize}}

\newtheorem{hej}{Theorem}[section]
\newtheorem{tjo}[hej]{Lemma}

\def\Pf{\noindent \textbf{Proof. }}
\def\End{\mathrm{End\,}}

\def\dim{\mathrm{dim\,}}

\def\der'{\mathfrak{der}'\,}
\def\der{\mathfrak{der}\,}
\def\str'{\mathfrak{str}'\,}
\def\str{\mathfrak{str}\,}

\def\C{\mathbb{C}}

\def\su{\mathfrak{su}}
\def\sp{\mathfrak{sp}}

\def\qed{\hspace{\stretch{1}} $\square$ \\
\noindent}

\newcommand{\al}{\alpha}
\newcommand{\de}{\delta}

\RequirePackage{hyperref}

\numberwithin{equation}{section}

\begin{document}

\hfill{\tt AEI-2009-046}\\
\vskip-10pt
\hfill {\tt \today}

 \pagestyle{empty}

\begin{center}

\vspace*{2cm}

\noindent
\begin{center}
{\LARGE \textsf{\textbf{Three-algebras, triple systems}} }\\
\vspace{.2cm}\hspace{.05cm}
{\LARGE \textsf{\textbf{\,and 3-graded Lie superalgebras}} }

\vskip 2truecm

{\large \textsf{\textbf{Jakob Palmkvist}}} \\
\vskip 1truecm
        {\ffam
        {Physique Th\'eorique et Math\'ematique\\
  Universit\'e Libre de Bruxelles \& International Solvay Institutes\\
  Boulevard du Triomphe, Campus Plaine, ULB-CP 231,\\BE-1050 Bruxelles, Belgium}\\[3mm]}
        {\tt jakob.palmkvist@ulb.ac.be} \\
\end{center}

\vskip 1cm

\centerline{\ffam\fser Abstract}
\end{center}

The three-algebras used by Bagger and Lambert in $N=6$ theories of ABJM type are in one-to-one correspondence with a certain type of Lie superalgebras.
We show that the description of three-algebras as generalized Jordan triple systems naturally leads to this correspondence. Furthermore, we show that simple three-algebras correspond to simple Lie superalgebras, and vice versa.
This gives a classification of simple three-algebras from the well known classification of simple Lie superalgebras.

\newpage

\pagestyle{plain}

\section{Introduction}
Superconformal Chern-Simons theories in three dimensions \cite{Schwarz:2004yj} have recently attracted much interest. 
Especially 
two different approaches have appeared in the literature, based on so called three-algebras and Lie superalgebras, respectively. In this paper we will connect these two algebraic structures to each other using a third one, {\it generalized Jordan triple systems}.

Bagger and Lambert proposed a model for multiple M2-branes with $N=8$ supersymmetry based on an algebraic structure called \textit{three-algebra}
\cite{Bagger:2006sk,Bagger:2007jr,Bagger:2007vi}.
The closure of the supersymmetry algebra was first shown by Gustavsson \cite{Gustavsson:2007vu} using a different (but equivalent \cite{Bagger:2007vi}) algebraic approach. 
It was later proven \cite{Papadopoulos:2008sk,Gauntlett:2008uf} that there is only one such simple three-algebra, leading to an $SO(4)$ gauge symmetry. 
Bagger and Lambert subsequently generalized 
their model
to one with $N=6$ supersymmetry,
based on a new, more general, notion of three-algebras \cite{Bagger:2008se}. 
It was shown 
that an infinite class of such three-algebras, parameterized by an integer $n\geq2$, gives the $N=6$ theories of Aharony, Bergman, Jafferis and Maldacena (ABJM) with gauge groups 
$SU(n) \times SU(n)$ \cite{Aharony:2008ug,Benna:2008zy,Aharony:2008gk}. For $n=2$ the supersymmetry enhances from $N=6$ to $N=8$, and the ABJM theory coincides with the original Bagger-Lambert-Gustavsson theory, in accordance with the isomorphism
$SO(4)=SU(2) \times SU(2)$. 
Schnabl and Tachikawa have 
classified all $N=6$ theories that can be obtained by such an enhancement from theories with a smaller amount of supersymmetry 
\cite{Schnabl:2008wj}. 

Three-dimensional conformal Chern-Simons theories with $N=4$ supersymmetry were studied by Witten and Gaiotto in relation to Janus configurations of four-dimensional $N=4$ super Yang-Mills theory \cite{Gaiotto:2008sd}. They showed that these novel theories can be classified in terms of Lie superalgebras. 
Hosomichi, Lee, Lee, Lee and Park extended the theories by Gaiotto and Witten by additional twisted hypermultiplets \cite{Hosomichi:2008jd}. In this way, they obtained the ABJM theory with
$U(m) \times U(n)$ gauge symmetry
as corresponding to the Lie supergroup $U(m|n)$ \cite{Hosomichi:2008jb}. 
The supersymmetry was thus enhanced from $N=4$ to $N=6$ in this case. Such an enhancement was shown also for another theory, corresponding to the Lie supergroup
$OSp(2|2n)$.

Generalized Jordan triple systems
are algebraic structures that on the one hand include three-algebras as special cases, and on the other hand have a well studied connection to Lie algebras
\cite{Kantor3.5}. 
Their relevance for the ABJM theory were first pointed out in 
\cite{Nilsson:2008kq}, where it was shown that instead of a three-algebra, the theory can equally well be formulated in terms of the associated Lie algebra. It was also shown that the associated Lie algebra is infinite-dimensional, but not of Kac-Moody type (thus it should not be confused with the Lie algebra of the gauge group.)
 
The main observation in the present paper is that the construction by Kantor \cite{Kantor3.5} of the associated Lie algebra in a simple way can be modified to the construction of an associated Lie \textit{superalgebra},
without (unlike other methods to construct Lie superalgebras from triple systems
\cite{Bars:1978yx,Kamiya1,Kamiya2,Kamiya3}) generalizing the 
definition of the triple system itself. We will show that 
when the generalized Jordan triple system is a three-algebra, the associated Lie superalgebra is finite-dimensional and 3-graded (with respect to a $\mathbb{Z}$-grading consistent with the usual $\mathbb{Z}_2$-grading). Conversely, any 3-graded Lie superalgebra gives rise to a generalized Jordan triple system which under certain conditions is a three-algebra. This naturally leads to
a one-to-one correspondence between three-algebras and this type of 
Lie superalgebras that we will describe in detail. 

The correspondence between three-algebras and 
Lie superalgebras has also been studied by 
Figueroa-O'Farrill et al.~\cite{deMedeiros:2008zh} in the context of the more general Cherkis-S\"amann three-algebras \cite{Cherkis:2008qr}. 
This description does not involve generalized Jordan triple systems, but is a special case of a general construction by Faulkner \cite{Faulkner2}.
One advantage of our method is that it enables us to refine the result: we will show that {simple} three-algebras correspond to {simple} Lie superalgebras and vice versa. (However, this can shown also with the method in \cite{deMedeiros:2008zh}. See note added below.) The simple Lie superalgebras that appear have been classified by Kac \cite{Kac77A,Kac77B} as $C(n+1)$ and $A(m,\,n)$.
Thus we arrive at a classification of the simple three-algebras defined by Bagger and Lambert in \cite{Bagger:2008se}.
In particular, 
the uniqueness of the three-algebra in \cite{Bagger:2006sk,Bagger:2007jr}
is from this point of view just a consequence of the well known classification by Kac of simple Lie superalgebras.

The paper is organized as follows. In section 2 we give our definition of a three-algebra (which is the same as in \cite{Bagger:2008se}) and show that it is a special case of a generalized Jordan triple system. We review how an arbitrary generalized Jordan triple system gives rise to an associated graded Lie algebra. Then we show how this construction can be modified to give a Lie superalgebra instead of a Lie algebra. This leads to Theorem 2.1, which says that there is a one-to-one correspondence between simple three-algebras and a certain type of simple Lie superalgebras. In section 3, we go in the opposite direction: we start with the Lie superalgebras $C(n+1)$ and $A(m,\,n)$, and describe how the corresponding triple systems can be obtained. Section 4 contains our conclusions, and the proof of the theorem is completed in an appendix.
\noindent
\subsubsection*{Note added}
After the preprint of this paper appeared, it was shown 
in \cite{FigueroaO'Farrill:2009pa}
that the refined result (that {simple} three-algebras correspond to {simple} Lie superalgebras and vice versa) 
also can be recovered from the construction in \cite{deMedeiros:2008zh}. See also \cite{deMedeiros:2009eq}.

\section{Three-algebras and triple systems}

The original definition of a three-algebra given by Bagger and Lambert \cite{Bagger:2007jr} has subsequently been generalized by many authors, among them Bagger and Lambert themselves \cite{Bagger:2008se}. We will in this paper use the two definitions in
\cite{Bagger:2007jr} and \cite{Bagger:2008se}, but in order to distinguish between them, we will talk about
`$N=6$ three-algebras' and `$N=8$ three-algebras', respectively.
When we only say `three-algebra' we mean an `$N=6$ three-algebra', since we are mainly interested in this more general case.

\subsection{Basic definitions}

An $N=6$ {\bf three-algebra} is a finite-dimensional complex vector space $V$ with a triple product $f : V \times V \times V \to V$ and an `inner product' $h: V \times V \to \C$, such that
\begin{itemize}
\item the triple product $(xyz) \equiv f(x,\,y,\,z)$ 
is linear in $x$ and $z$ but antilinear in $y$:
\begin{align}
\al (xyz)=((\al x)yz)=(x(\al^\ast y)z)=(xy(\al z))
\end{align}
for any complex number $\alpha$ (where $\ast$ is the complex conjugate),
\item the triple product 
satisfies the identities
\begin{align}
(uv(xyz))-(xy(uvz))&=((uvx)yz)-(x(vuy)z),\label{N=6ta1}\\
(xyz)&=-(zyx)\label{N=6ta2},
\end{align}
\item the inner product $\la x | y \ra \equiv h(x,\,y)$ is linear in $x$ and antilinear in $y$,
\item the inner product satisfies the identities
\begin{align}
\la w | (xyz)\ra=\la y | (zwx) \ra&=\la(wzy)|x\ra=\la(yxw)|z\ra, 
\label{N=6ta3}\\
\la x |y\ra&=\la y |x\ra^{\ast}, \label{N=6ta4}
\end{align}
\item the inner product is positive-definite.
\end{itemize}

An $N=8$ {\bf three-algebra} is an $N=6$ three-algebra $V$ with a \textbf{conjugation} $C$ (an antilinear involution) such that 
the triple product satisfies $(xC(y)z)=-(yC(x)z)$ and the inner product $h$ is real. This implies that the triple product is totally antisymmetric and that the inner product is symmetric. Then the identity (\ref{N=6ta1}) takes the form of a \textit{Leibniz rule},
\begin{align}
(uv(xyz))=((uvx)yz)+(x(uvy)z)+(xy(uvz)).
\end{align}
In this form it is often called the {\it fundamental identity}. Moreover, for an $N=8$ three-algebra (\ref{N=6ta3}) expresses an \textit{invariance} of the inner product,
\begin{align}
\la (xyw) | z \ra + \la w | (xyz) \ra = 0.
\end{align}

Let $V$ be a 
$N=6$ three-algebra, and let $T^{a}$ be an orthonormal basis of $V$, for $a=1,\,2,\,\ldots,\,\dim{V}$. 
In analogy with Lie algebras we introduce \textbf{structure constants} 
$f^{a}{}_{b}{}^{c}{}_d$ for $V$, which specify the triple product by
\begin{align}
(T^a T^b T^c) = f^{a}{}_{b}{}^{c}{}_d T^d.
\end{align}
We will see that it is natural to put the second index downstairs. This is related to the fact that the triple product is not linear but antilinear in the second argument.
The identities (\ref{N=6ta1})--(\ref{N=6ta3}) can now be written
\begin{align}
f^{a}{}_{b}{}^{c}{}_d f^{e}{}_{f}{}^{d}{}_g-
f^{e}{}_{f}{}^{c}{}_d f^{a}{}_{b}{}^{d}{}_g&=
f^{e}{}_{f}{}^{a}{}_d f^{d}{}_{b}{}^{c}{}_g-
(f^{f}{}_{e}{}^{b}{}_d{})^\ast f^{a}{}_{d}{}^{c}{}_g,\label{N=6ta1ny}\\
f^{a}{}_{b}{}^{c}{}_d&=-f^{c}{}_{b}{}^{a}{}_d, \label{N=6ta2ny}\\ 
(f^{a}{}_{b}{}^{c}{}_{d})^\ast=(f^{c}{}_{d}{}^{a}{}_{b})^\ast&=
f^{b}{}_{a}{}^{d}{}_{c}=f^{d}{}_{c}{}^{b}{}_{a}.\label{N=6ta3ny}
\end{align}
This means that 
$f$ is 
antisymmetric not only in the upper pair of indices, but also in the lower pair.
In \cite{Nilsson:2008kq} the identity (\ref{N=6ta3ny}) was used to rewrite (\ref{N=6ta1ny}) as
\begin{align}
f^{a}{}_g{}^{[b}{}_d f^{c]}{}_e{}^g{}_f=
f^{b}{}_g{}^{c}{}_{[e} f^{a}{}_{f]}{}^g{}_d.
\end{align}
We follow the notation in \cite{Nilsson:2008kq}, which means that we put the indices of $f$ 
in a different order compared to Bagger and Lambert \cite{Bagger:2008se}: 
\begin{align}
(f^{a}{}_{b}{}^{c}{}_d)_{\text{NP}}&=(f^{ac\bar b}{}_d)_{\text{BL}}.
\end{align}
The identities (\ref{N=6ta1ny}) and (\ref{N=6ta3ny}) should be compared to (2) and (4), respectively, in 
\cite{Bagger:2008se}.

A difference compared to \cite{Nilsson:2008kq} is that $f$ there was linear in all three arguments, also the second one. But as well as $f$ and $h$ given here, we can always 
consider the triple product $\tilde f$ and the inner product $\tilde h$ in $V$, given by
\begin{align}
\tilde f (x,\,y,\,z)&=f(x,\,C(y),\,z), & \tilde h (x,\,y)&=h(x,\,C(y))
\end{align}
where $C$ is a {conjugation}. 
Thus $\tilde f$ and $\tilde h$ are linear in all arguments, and we can consider the $N=6$ three-algebra $V$ as a special kind of a {\bf generalized Jordan triple system} \cite{Kantor3.5}. By definition, this is a vector space with a {\it trilinear} triple product that satisfies the identity (\ref{N=6ta1}). 

A {\bf weak ideal} of a generalized Jordan triple system $V$ is a subspace $W$ such that
$(VVW) \subseteq W$ 
and $(WVV) \subseteq W$. For an $N=6$ three-algebra these two conditions are equivalent, because of the antisymmetry of the triple product. If in addition $(VWV) \subseteq W$ then $W$ is called an {\bf ideal} of $V$. Thus any ideal is a weak ideal. But for $N=6$ three-algebras also the converse is true, as we will show in the appendix. 

The triple system $V$ is {\bf simple} if the only ideals of $V$ are $0$ and $V$ itself. If it is not simple then we can 
`make it simple' by factoring out the maximal ideal.

\subsection{The associated graded Lie algebra}

We will now describe how a generalized Jordan triple system $V$ gives rise to a graded Lie algebra \cite{Kantor3.5} in the special case when $V$ is an $N=6$ three-algebra. This was already done in \cite{Nilsson:2008kq}, but the description here will be more detailed.
Also in the general case, our approach is somewhat different from the original one by Kantor (but equivalent).

Any vector space $V$ gives rise to a graded Lie algebra
\begin{align}
\tilde U&=\tilde{U}{}_{-1}\oplus\tilde U_0\oplus\tilde U_1\oplus\tilde U_2\oplus\cdots\nn\\
&=\tilde{U}{}_{-1}\oplus\tilde U_0\oplus\tilde U_+,
\end{align}
which is the direct sum of subspaces $\tilde U_k$ for all integers $k$,
such that
$[\tilde U_i,\,\tilde U_j] \subseteq \tilde U_{i+j}$, and 
$\tilde U_k = 0$ for $k\leq-2$ \cite{Kantor-graded}.
The subspaces $\tilde U_k$ for $k\geq-1$ are defined recursively, starting with $\tilde U_{-1}=V$.
For each $k\geq0$, the subspace $\tilde U_{k}$ is then defined as the vector space of all linear maps $V \to \tilde U_{k-1}$.
Thus $\tilde U_0=\End{V}$, and $\tilde U_1$ consists of linear maps $V \to \End{V}$.
The Lie bracket is defined recursively by 
\begin{align} \label{recdefbracket}
[a,\,b]=(\text{ad }{a}) \circ b - (\text{ad }{b}) \circ a,
\end{align}
starting from $[u,\,v]=0$ for $u,\,v \in \tilde{U}_{-1}$. 
Here any element $u$ in $\tilde U_{-1}=V$ should be considered as a constant map $V \to V$, given by $u(v)=u$ for all $v \in V$.
The Jacobi identity can be shown by induction.

Suppose now that $V$ is an $N=6$ three-algebra with an orthonormal basis $T^a$ and structure constants $f^a{}_b{}^c{}_d$.
Define the elements $S^a{}_b \in \tilde U_0$ 
and $\bar T_a \in \tilde U_{1}$ by
\begin{align}
S^a{}_b\ : \qquad\quad V &\to \tilde U_{-1}=V, & S^a{}_b(T^c)&=f^a{}_b{}^c{}_d T^d,\nn\\
\bar T_a\ :\qquad\quad V &\to \tilde{U}_0, & \bar T_a (T^b) &= S^b{}_a.
\end{align}
Set $U_{-1}=\tilde U_{-1}=V$ and let $U_0$ and $U_1$ be the subspaces of 
$\tilde U_0$ and $\tilde U_1$ spanned by all elements $S^a{}_b$
and $\bar T_a$, respectively.
Furthermore, let $U_+ = U_{1} \oplus U_{2}\oplus\cdots$ be the subalgebra of 
$\tilde U_+=\tilde U_1
\oplus \tilde U_2 \oplus\cdots$ generated by $U_1$ (with grading inherited from $\tilde U_+$).

Since the inner product in an $N=6$ three-algebra is positive-definite and thus non-degenerate, it follows from (\ref{N=6ta3}) that 
if $(xuy)=(xvy)$ for all $x$ and $y$, then $u=v$.
This means that $U_1$ is isomorphic to $U_{-1}$.
The map 
\begin{align} \label{tau1}
\tau \ :\  \bar T_a \mapsto T^a
\end{align}
between the bases of $U_1$ and $U_{-1}$ is one-to-one and
can be extended by linearity to an isomorphism between the 
vector spaces $U_1$ and $U_{-1}$. However, here we choose to extend it by \textit{antilinearity}
to an \textit{anti-isomorphism}
$\tau: U_+ \to U_-$, 
where, for each $k\geq 2$, we introduce  
a vector space $U_{-k}$ isomorphic to $U_k$,
and let $U_-$ be the direct sum 
$U_- = U_{-1} \oplus U_{-2}\oplus\cdots$
of all these vector spaces. 
Furthermore, we can define a Lie algebra 
structure
on $U_-$ by
\begin{align}
[\tau(a),\,\tau(b)] \equiv \tau([a,\,b]).
\end{align}
Then $\tau$ extends to a {Lie algebra} anti-isomorphism from $U_{+}$ to $U_-$. These two isomorphic Lie algebras are generated by $U_{1}$ and $U_{-1}$, respectively. We can define a Lie algebra structure on the direct sum
\begin{align}
U&=\cdots \oplus U_{-2}\oplus U_{-1} \oplus U_0 \oplus U_1\oplus U_2 \oplus \cdots\nn\\
&=U_-\oplus U_0 \oplus U_+
\end{align}
by combining the Lie algebra structures on the 
subspaces $U_-$ and $U_{-1}+U_0+U_+$. 
It follows by construction that the commutation relations 
involving $U_{\pm1}$ and $U_0$ are
\begin{align}
[S^a{}_b,\,T^c]&=f^a{}_b{}^c{}_dT^d,\nn\\
{[}S^a{}_b,\,\bar T_d]&=-f^a{}_b{}^c{}_d\bar T_c,\nn\\
{[}\bar T_b,\,T^a]&=S^a{}_b,\nn\\
[S^a{}_b,\,S^c{}_d]&=f^a{}_b{}^c{}_e S^e{}_d -
f^a{}_b{}^e{}_d S^c{}_e.
\label{kommrelgraded}
\end{align}
All other commutation relations then follow from 
(\ref{kommrelgraded}) by the Jacobi identity and the fact that $U$ is generated by $T^a$ and $\bar T_ a$. It follows from (\ref{kommrelgraded}) that $\tau$ can be extended to a conjugation on $U$ by 
the inverse of (\ref{tau1}),
\begin{align}
\tau\ :\ T^a \mapsto  \bar T_a.
\end{align}
This conjugation acts on $U_0$ in the following way,
\begin{align}
S^a{}_b = [\bar T_b,\,T^a]\mapsto [T^b,\,\bar T_a]=-[\bar T_a,\,T^b]=-S^b{}_a.
\end{align}
All elements $\Lambda$ in $U_0$ that are invariant under this conjugation constitute  
a real form of the complex Lie algebra $U_0$.
If we write $\Lambda = \Lambda_a{}^bS^a{}_b$ this means 
$(\Lambda_a{}^b)^\ast=-\Lambda_b{}^a$.
It also follows from (\ref{N=6ta1}) that they act as {\it derivations} on $V$:
\begin{align}
\Lambda((xyz))=(\Lambda(x)yz)+(x\Lambda(y)z)+(xy\Lambda(z)).
\end{align}
In \cite{Nilsson:2008kq} we studied the properties of the graded Lie algebra 
$U$ associated to a three-algebra $V$. We found that it is infinite-dimensional, but not of Kac-Moody type. 
Here we will instead modify the associated Lie algebra to a Lie 
{\it superalgebra}.

\subsection{The associated graded Lie superalgebra}
In the same way as $V$ gives rise to a graded Lie algebra $U$, it also gives rise to a graded Lie {superalgebra} $\bfU$. 
We recall that a \textbf{Lie superalgebra} \cite{Kac77A,Kac77B} is a $\mathbb{Z}_2$-graded algebra $\vG=\vG_{(0)} \oplus \vG_{(1)}$ with a \textbf{Lie superbracket} such that
\begin{align}
\dlb a,\,b \drb &= -(-1)^{pq}\dlb b,\,a \drb \in \vG_{(p+q)}
\end{align}
if $a \in \vG_{(p)}$ and $b \in \vG_{(q)}$. We say that $p$ and $q$ are the \textbf{degrees} of $a$ and $b$, respectively.
The degrees should be considered as elements in the field $\mathbb{Z}_2 =\{0,\,1\}$ so that $1+1=0$. If 
$pq=0$, we write $\dlb a,\,b \drb=[a,\,b]$,
and if $pq=1$, we write $\dlb a,\,b \drb=\{a,\,b\}$.
Furthermore the superbracket is required to satisfy the \textbf{Jacobi superidentity}
\begin{align}
(-1)^{pr} \dlb \dlb a,\,b \drb,\,c\drb + (-1)^{qp} \dlb \dlb b,\,c \drb,\,a\drb + (-1)^{rq} \dlb \dlb c,\,a \drb,\,b\drb = 0,
\end{align}
where $p,\,q,\,r$ are the degrees of $a,\,b,\,c$, respectively.
For more information about Lie superalgebras, see for example \cite{Frappat}.

A Lie superalgebra is already a $\mathbb{Z}_2$-graded algebra, but when we talk about graded algebras here, we refer to $\mathbb{Z}$-gradings. In particular, when we talk about a graded Lie superalgebra 
\begin{align}
\vG=\cdots\oplus\vG_{-1}\oplus\vG_{0}\oplus\vG_{1}\oplus\cdots,
\end{align}
we refer to a $\mathbb{Z}$-grading that is {\bf consistent} 
with the $\mathbb{Z}_2$-grading.
This means that
$\vG_{(0)}$ is the sum of all $\vG_k$ for $k$ even, while $\vG_{(1)}$ is the sum of all $\vG_k$ for $k$ odd. 

For any vector space $V$, we can now define a graded Lie superalgebra $\tilde\bfU$ as the same vector space as $\tilde U$, 
defined in the preceding subsection.
Thus $\tilde \bfU$ is a direct sum of subspaces 
$\tilde \bfU_k=\tilde U_k$ for $k\geq-1$. We equip $\tilde \bfU$ with a Lie superalgebra structure by modifying the recursively defined Lie bracket
(\ref{recdefbracket}) to
\begin{align}\label{recdefsuperbracket}
{[\![}a,\,b{]\!]}=(\text{ad }{a}) \circ b - (-1)^{pq} (\text{ad }{b}) \circ a,
\end{align}
where $p$ and $q$ are the degrees of $a$ and $b$, respectively.
The Jacobi {superidentity} can then be shown by induction. 

If $V$ is a generalized Jordan triple system, we set $\bfU_{-1}=\tilde \bfU_{-1}=V$ and let $\bfU_0$ and $\bfU_1$ be the subspaces of $\tilde \bfU_0$ and $\tilde \bfU_1$ spanned by all elements $S^a{}_b$
and $\bar T_a$, respectively, in analogy with the construction of the Lie algebra $U$.
Thus $\bfU_{\pm1}=U_{\pm1}$ and $\bfU_0=U_0$. But when we define $\bfU_+$ as the subalgebra of $\tilde\bfU_+=\tilde\bfU_1\oplus\tilde\bfU_2\oplus\cdots$ 
generated by $\bfU_1$, and let $V$ be an $N=6$ three-algebra, a difference arises compared to $U_+$, as we will see now.
The subspace $\bfU_2=\{\bfU_1,\,\bfU_1\}$ is spanned by elements
$\{\bar T_a,\,\bar T_b\}$ which are maps from $V$ to $\bfU_1$. It follows from 
(\ref{recdefsuperbracket}) that they act as
\begin{align}
\{\bar T_a,\,\bar T_b\}(T^c)&=[ \bar T_a,\,\bar T_b(T^c)]
+[ \bar T_b,\,\bar T_a(T^c)]\nn\\
&=[ T_a,\,S^c{}_b]
+[ T_b,\,S^c{}_a]\nn\\
&=f^c{}_b{}^d{}_a{}\bar T_d+f^c{}_a{}^d{}_b{}\bar T_d,
\end{align}
and this is zero for an $N=6$ three-algebra.
Since $\bfU_{+}$ is generated by $\bfU_1$, this means that $\bfU_+=\bfU_1$, and the Lie superalgebra is in fact 3-graded,
\begin{align}
\bfU=\bfU_{-1}\oplus\bfU_0\oplus\bfU_1.
\end{align}
We note that the same happens for the Lie algebra associated to a {\it Jordan triple system}, which is a generalized Jordan triple system such that $(xyz)=(zyx)$. It follows from this symmetry of the triple product that the subspaces $U_{\pm2}$ vanish, in the same way as $\bfU_{\pm2}$ do for an $N=6$ three-algebra.
The complete set of commutation relations in $\bfU$ is then given by
\begin{align}
\dlb S^a{}_b,\,T^c\drb&=[S^a{}_b,\,T^c]=f^a{}_b{}^c{}_dT^d,\nn\\
\dlb S^a{}_b,\,\bar T_d \drb &={[}S^a{}_b,\,\bar T_d]=-f^a{}_b{}^c{}_d\bar T_c,
\nn\\
\dlb\bar T_b,\,T^a\drb&=\{\bar T_b,\,T^a\}=S^a{}_b,\nn\\
\dlb S^a{}_b,\,S^c{}_d \drb &=[S^a{}_b,\,S^c{}_d]=f^a{}_b{}^c{}_e S^e{}_d -
f^a{}_b{}^e{}_d S^c{}_e.
\label{kommrelgradedsuper-ny2}
\end{align}
Next we want to extend the anti-isomorphism
\begin{align}
\tau : \bfU_1 \to \bfU_{-1} ,\quad \bar T_a \mapsto T^a
\end{align}
to the whole of $\bfU$ in a way such that $\tau(S^a{}_b)=-S^b{}_a$ as before. But then we cannot extend it by its inverse as we did for the Lie algebra $U$. Instead we have to take
\begin{align}
\tau : \bfU_{-1} \to \bfU_1 ,\quad  T^a \mapsto -\bar T_a. 
\end{align}
Then we get
\begin{align}
S^a{}_b = \{\bar T_b,\,T^a\}\mapsto -\{T^b,\,\bar T_a\}=-\{\bar T_a,\,T^b\}=-S^b{}_a.
\end{align}
The resulting anti-automorphism satisfies
\begin{align}
[\tau(a),\,\tau(b)]&=\tau([a,\,b]), & \tau^2(c)=(-1)^k(c) \label{superkonjvillkor}
\end{align}
for all $a,\,b\in \bfU$ and $c\in \bfU_k$. We call any such anti-automorphism $\tau$ on a graded 
Lie superalgebra (with consistent grading)
\begin{align}
\vG &= \cdots \oplus \vG_{-1}\oplus\vG_0\oplus\vG_1\oplus\cdots,
\end{align}
such that $\tau(\vG_k)=\vG_{-k}$ for all integers $k$ and (\ref{superkonjvillkor}) 
holds for all $c\in \vG_k$, a {\bf graded superconjugation}.

Before stating the main result of this paper, Theorem 2.1 below, we need one more definition.
A bilinear form $\kappa$ on a Lie superalgebra $\vG=\vG_{(0)}\oplus\vG_{(1)}$ is said to be \textbf{consistent} and \textbf{supersymmetric} if 
\begin{align}
\kappa(a,\,b)&=\tfrac12\big((-1)^p+(-1)^q\big)\,\kappa(b,\,a),\nn
\end{align}
for any $x \in \vG_{(p)}$ and $y \in \vG_{(q)}$ (where $p,\,q=0,\,1$).
If it furthermore is invariant,
\begin{align}
\kappa(\dlb a,\,b\drb,\,c)&=\kappa(a,\,\dlb b,\,c\drb).
\end{align}
for any $a,\,b,\,c \in \vG$,
then we call it an
\textbf{inner product}. 
\begin{hej}
There is a one-to-one correspondence between 
simple $N=6$ three-algebras $V$ and simple 3-graded Lie superalgebras 
$\bfU$ with an inner product $\kappa$ 
and a graded superconjugation $\tau$ such that
\begin{align} \label{kappatauvillkor}
\kappa(a,\,\tau(b))&=\kappa(b,\,\tau(a))^\ast, & \kappa(a,\,\tau(a))&>0,
\end{align}
for all $a,\,b \in \bfU$. 
\end{hej}
\Pf We have already described how $L(V)\equiv\bfU$ is defined from $V$. 
The inner product $\kappa$ and the graded superconjugation $\tau$ are then defined by
\begin{align}
\tau(T^a)&=-\bar T_a, & \tau(\bar T_a)&=T^a, & \tau(S^a{}_b)&=-S^b{}_a,\nn
\end{align}
\begin{align}
\kappa(T^a,\,\bar T_b)&=-\kappa(\bar T_b,\,T^a)=\delta^a{}_b, &
\kappa(S^a{}_b,\,S^c{}_d)&=f^a{}_b{}^c{}_d.
\end{align}
It is straightforward to show that they satisfy the requirements in the theorem.

Conversely, start with the 3-graded Lie superalgebra $\bfU$. Then we take $\bfU_{-1}$ to be the $N=6$ three-algebra $K(\bfU)\equiv V$ with the triple product
\begin{align} \label{trippelprodfransla}
(xyz)=\dlb \dlb x,\,\tau(y)\drb,\,z\drb = [ \{ x,\,\tau(y)\},\,z ]
\end{align}
and the inner product
\begin{align} \label{innerprodfransla}
\la x | y \ra = \kappa(x,\,\tau(y)).
\end{align}
Again, it is straightforward to show that the definition of an $N=6$ three-algebra is satisfied. It is also easy to see that the map $L$ is invertible: 
$K(L(V))=V$ for any $N=6$ three-algebra $V$.

To prove that $L(K(\bfU))=\bfU$ for any 3-graded Lie superalgebra $\bfU$ that satisfies the requirements in the theorem, there are two non-trivial conditions that we have to show. Firstly, that 
$\{\bfU_{-1},\,\bfU_1\}=\bfU_0$, and secondly, that $c=0$ whenever $c\in \bfU_{0}$ and $[c,\,x]=0$ for all 
$x\in \bfU_{-1}$.
We note that the subspace
\begin{align}
\bfU_{-1} + \{\bfU_{-1},\,\bfU_1\} + \bfU_{1}
\end{align}
is an ideal of $\bfU$, and since $\bfU$ is simple, we must have
$\{\bfU_{-1},\,\bfU_1\}=\bfU_0$. Then also the second condition follows, since otherwise all such elements $c$ would form an ideal.

It remains to show that $L(V)$ is a simple Lie superalgebra if and only if $V$ is a simple $N=6$ three-algebra. We will do this in the appendix.
\qed

\noindent
It follows that a classification of all simple three-algebras is equivalent to the classification of all simple Lie superalgebras that satisfy the conditions in the theorem.
The simple 3-graded Lie superalgebras have been classified by Kac \cite{Kac77A,Kac77B}, and they are called \textit{classical Lie superalgebras of type I}. They are denoted $A(0,\,n)$, $A(m,\,n)$, $C(n+1)$ and $P(n+1)$ for all integers $m,\,n\geq 1$. Among these, $P(n+1)$ is a \textit{strange} Lie superalgebra, which in particular means that 
$\dim{\bfU_{-1}} \neq \dim{\bfU_{1}}$ in the 3-grading
$P(n)=\bfU_{-1}+\bfU_0+\bfU_1$. Thus $P(n+1)$ does not admit a graded superconjugation. For the remaining \textit{basic} Lie superalgebras $A(0,\,n)$, $A(m,\,n)$ and $C(n+1)$ there is a unique 
3-grading up to automorphisms. Furthermore, there is
an inner product $\kappa$ and a graded superconjugation $\tau$ such that the requirements are satisfied. 

It is a general result that an inner product on a simple Lie superalgebra is unique up to an overall constant, so we only have to show that there is a superconjugation $\tau$ that is compatible with this inner product. The inner product reduces to the ordinary Killing form on the subalgebra $\bfU_0$ (which is a semisimple Lie algebra), and the superconjugation $\tau$ reduces to an ordinary conjugation on $\bfU_0$. This means that it defines a real form of the complex Lie algebra $\bfU_0$, consisting of all elements $x + \tau(x)$, where $x \in \bfU_0$. Then the requirement that $\kappa$ be positive-definite on this real Lie algebra defines $\tau$ uniquely on $\bfU_0$, up to automorphisms.
In the next section we will see that there 
is even a way to extend $\tau$ to the $\bfU_{\pm1}$ subspaces such that the conditions
(\ref{kappatauvillkor}) are satisfied on the full Lie superalgebra $\bfU$.

\section{From Lie superalgebras back to three-algebras}

We will now change perspective and focus on the simple Lie superalgebras 
$C(n+1)$ and $A(m,\,n)$ that satisfy the conditions in the theorem.
Physicists are perhaps more familiar with the notation $\sp(2|2n,\,\C)$, $\su(m+1|n+1,\,\C)$ and $\mathfrak{psu}(n+1|n+1,\,\C)$ for $C(n+1)$, $A(m,\,n)$ and $A(n,\,n)$, respectively, where $m\neq n$.
We will review how these Lie superalgebras
can be constructed from their Cartan matrices or Dynkin diagrams. The Cartan matrices and Dynkin diagrams that we use are always the \textit{distinguished} ones. For more details, we refer to \cite{Kac77A,Kac77B,Frappat}. 

We will define an inner product $\kappa$ and superconjugation $\tau$ on the Lie superalgebras $C(n+1)$ and $A(m,\,n)$ that satisfy the conditions (\ref{kappatauvillkor}), whereupon the construction of the associated $N=6$ three-algebra follows by (\ref{trippelprodfransla}) and (\ref{innerprodfransla}). Repeated indices should not be summed over in this section.

\subsection{The Lie superalgebras $C(n+1)$}

For any integer $n \geq 1$, the Lie superalgebra $C(n+1)$
is given by the entries $A_{ij}$ of the 
Cartan matrix
\begin{align} \label{ccartan}
&\begin{array}{rrrrrrr}
\qquad
\text{\scriptsize $0$}&
\  \,\,\,
\text{\scriptsize $1$}&
\  \,\,\,\,
\text{\scriptsize $2$}&
&\qquad&\qquad\quad\ \ 
\text{\scriptsize $n$}
\end{array}\nn\\
\begin{array}{r}
\text{\scriptsize $0$}\\
\text{\scriptsize $1$}\\
\text{\scriptsize $2$}\\
\phantom{\vdots}\\
\text{\scriptsize $n-2$}\\
\text{\scriptsize $n-1$}\\
\text{\scriptsize $n$}
\end{array}
&\left(
\begin{array}{r|rrrrrr}
0&-1& 0 &  \cdots & 0 & 0 & 0\\ \hline
 -1 & 2 & -1 &  \cdots & 0 & 0 & 0\\
 0 & -1 & 2 &  \cdots & 0 & 0 & 0\\
\vdots\, &\vdots\, &\vdots\, & \ddots &\vdots\, &\vdots\, &\vdots\,\\           
0&0&0& \cdots & 2 & -1 & 0\\
0&0&0& \cdots & -1 & 2 & -2\\
0&0&0& \cdots &0 & -1 & 2                 
\end{array}
\right)\\
&\qquad\quad\ \, \underbrace{
\qquad\qquad\qquad\qquad\qquad\qquad}_{n}\nn
\end{align}
or, equivalently, by the Dynkin diagram
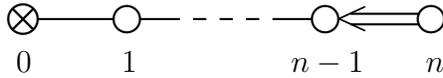
\begin{center}
\begin{picture}(440,30)(20,0)
\put(163,-10){$0$}
\put(203,-10){$1$}
\put(267,-10){$n-1$}
\put(318,-10){$n$}
\thicklines
\put(159,7){$\bigotimes$}
\multiput(280,10)(40,0){2}{\circle{10}}
\multiput(289.5,12)(40,0){1}{\line(1,0){26}}
\multiput(289.5,8)(40,0){1}{\line(1,0){26}}
\put(205,10){\circle{10}}
\multiput(171,10)(40,0){1}{\line(1,0){29}}
\multiput(210,10)(50,0){2}{\line(1,0){15}}
\multiput(230,10)(10,0){3}{\line(1,0){5}}
\put(283.5,5.7){\Large$<$}
\end{picture}
\end{center}
\vspace*{0.8cm}
To each row and column in the Cartan matrix, and to each node in the Dynkin diagram, we associate three \textbf{Chevalley generators} $e_i,\,f_i,\,h_i$
where $i=0,\,1,\,\ldots,\,n$ according to the labelling above.

The Lie superalgebra $C(n+1)$ is now generated by the Chevalley elements modulo the \textbf{Chevalley relations}
\begin{align}\label{chevrel}
\dlb e_p,\,f_q\drb&=
[e_p,\,f_q]=
\de_{pq}h_q,
&\dlb h_i,\,e_j\drb&=[h_i,\,e_j]=A_{ij}e_j,\nn\\
\dlb e_0,\,f_0\drb&=
\{e_0,\,f_0\}=
h_0
&\dlb h_i,\,f_j\drb &=[h_i,\,f_j]=-A_{ij}f_j,
\end{align}
for $(p,\,q)\neq(0,\,0)$,
and, for $i\neq j$, the \textbf{Serre relations}
\begin{align} \label{serrel}
(\text{ad }e_i)^{1+|A_{ij}|}(e_j)=(\text{ad }f_i)^{1+|A_{ij}|}(f_j)=0.
\end{align}
The subspace $\bfU_{-1}$ in the 3-grading $C(n+1)=\bfU_{-1}\oplus\bfU_0\oplus\bfU_1$ is spanned by all multiple commutators where the Chevalley generator $e_0$, corresponding to the 'grey' node above, appears exactly once. Likewise, 
$\bfU_{1}$ is spanned by all 
multiple commutators where $f_0$ appears exactly once. The subalgebra $\bfU_0$ is the direct sum of a one-dimensional Lie algebra and the simple complex Lie algebra $C_n$. The Dynkin diagram of $C_n$ is obtained by deleting the 
grey node, and its compact real form is $\sp(2n)$. 

Now we define the inner product $\kappa$ and the superconjugation $\tau$ for the Chevalley generators.
For the inner product we have
\begin{align}
\kappa(h_{n-1},\,h_n)&=1, &
\kappa(e_n,\,f_n)&=-\tfrac12      
\end{align}
and otherwise
\begin{align}
\kappa(h_i,\,h_j)&=-A_{ij}, &
\kappa(e_i,\,f_j)&=-\delta_{ij}.
\end{align}
The superconjugation is given by
\begin{align}
\tau(e_0)&=-f_0,&
\tau(e_p)&=-f_p,\nn\\
\tau(f_0)&=e_0,&
\tau(f_p)&=-e_p,\nn\\
\tau(h_0)&=-h_0,&
\tau(h_p)&=-h_p,
\end{align}
for $p>0$.
The inner product of two arbitrary elements then follows by invariance, and the action of the superconjugation on an arbitary element follows by the homomorphism property.

\subsection{The Lie superalgebras $A(m,\,n)$} 

For any integers $m,\,n \geq 0,\,(m,\,n)\neq(0,\,0)$, the Lie superalgebra 
$A(m,\,n)$
is given by the entries $A_{ij}$ of the 
Cartan matrix
\begin{align} \label{acartan}
&\begin{array}{rrrrrrrrrrrrr}
\quad\ \text{\scriptsize $-m$} &
\quad&
\quad\ &
\qquad&
\text{\scriptsize $-2$}&
\ 
\text{\scriptsize $-1$}&
\  \,\,
\text{\scriptsize $0$}&
\  \,\,\,\,
\text{\scriptsize $1$}&
\  \,\,\,
\text{\scriptsize $2$}&
&\qquad&\qquad\quad\ \ 
\text{\scriptsize $n$}
\end{array}\nn\\
\begin{array}{r}
\text{\scriptsize $-m$}\\
\text{\scriptsize $-m+1$}\\
\text{\scriptsize $-m+2$}\\
\phantom{\vdots}\\
\text{\scriptsize $-2$}\\
\text{\scriptsize $-1$}\\
\text{\scriptsize $0$}\\
\text{\scriptsize $-1$}\\
\text{\scriptsize $-2$}\\
\phantom{\vdots}\\
\text{\scriptsize $n-2$}\\
\text{\scriptsize $n-1$}\\
\text{\scriptsize $n$}
\end{array}
&\left(
\begin{array}{rrrrrr|r|rrrrrr}
-2 & 1 & 0 &  \cdots & 0 & 0 & 0&0&0& \cdots & 0 & 0 & 0\\
1 & -2 & 1 &  \cdots & 0 & 0 & 0&0&0& \cdots & 0 & 0 & 0\\
0 & 1 & -2 &  \cdots & 0 & 0 & 0&0&0& \cdots & 0 & 0 & 0\\
\vdots\, &\vdots\, &\vdots\, & \ddots &\vdots\, &\vdots\, &\vdots\,
&\vdots\, &\vdots\, &  &\vdots\, &\vdots\, &\vdots\,\\
0&0&0& \cdots & -2 & 1 & 0 & 0 & 0 & \cdots & 0 & 0 & 0\\ 
0&0&0& \cdots & 1 & -2 & 1 & 0 & 0 &  \cdots & 0 & 0 & 0\\ \hline
0&0&0& \cdots & 0 & 1 & 0 & -1 & 0 &  \cdots & 0 & 0 & 0\\ \hline
0&0&0& \cdots & 0 & 0 & -1 & 2 & -1 &  \cdots & 0 & 0 & 0\\
0&0&0& \cdots & 0 & 0 & 0 & -1 & 2 &  \cdots & 0 & 0 & 0\\
\vdots\, &\vdots\, &\vdots\, &  &\vdots\, &\vdots\, &\vdots\,
&\vdots\, &\vdots\, & \ddots &\vdots\, &\vdots\, &\vdots\,\\           
0&0&0& \cdots &0&0&0&0&0& \cdots & 2 & -1 & 0\\
0&0&0& \cdots &0&0&0&0&0& \cdots & -1 & 2 & -1\\
0&0&0& \cdots &0&0&0&0&0& \cdots &0 & -1 & 2                 
\end{array}
\right)\\
&\quad\ \underbrace{
\qquad\qquad\qquad\qquad\qquad\qquad}_{m}\qquad\quad
\underbrace{
\qquad\qquad\qquad\qquad\qquad\qquad}_{n}\nn
\end{align}
or, equivalently, by the Dynkin diagram
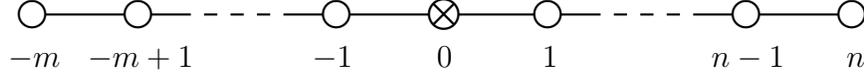
\begin{center}
\scalebox{1}{
\begin{picture}(335,30)
\put(1,-10){$-m$}
\put(31,-10){$-m+1$}
\put(116,-10){$-1$}
\put(163,-10){$0$}
\put(203,-10){$1$}
\put(267,-10){$n-1$}
\put(318,-10){$n$}
\thicklines
\multiput(10,10)(40,0){2}{\circle{10}}
\multiput(15,10)(40,0){1}{\line(1,0){30}}
\multiput(55,10)(50,0){2}{\line(1,0){15}}
\multiput(75,10)(10,0){3}{\line(1,0){5}}
\multiput(130,10)(40,0){1}{\line(1,0){30}}
\put(125,10){\circle{10}}
\put(159,7){$\bigotimes$}
\multiput(280,10)(40,0){2}{\circle{10}}
\multiput(285,10)(40,0){1}{\line(1,0){30}}
\put(205,10){\circle{10}}
\multiput(171,10)(40,0){1}{\line(1,0){29}}
\multiput(210,10)(50,0){2}{\line(1,0){15}}
\multiput(230,10)(10,0){3}{\line(1,0){5}}
\end{picture}}\end{center}
\vspace*{0.8cm}

The Lie superalgebra $A(m,\,n)$ can now be constructed from its Cartan matrix or its Dynkin diagram in the same way as $C(n+1)$ in the preceding subsection, starting from $3(m+n+1)$ {Chevalley generators} $e_i,\,f_i,\,h_i$,
where now $i=-m,\,-m+1,\,\ldots,\,n$ according to the labelling above.
The Lie superalgebra $A(m,\,n)$ is then generated by these elements modulo the Chevalley-Serre relations (\ref{chevrel})--(\ref{serrel})
and, if $m,\,n > 0$, the supplementary relations
\begin{align}
0&=(\text{ad }e_0)\circ(\text{ad }e_1)\circ(\text{ad }e_0)\circ(\text{ad }e_{-1})\nn\\&=
(\text{ad }e_0)\circ(\text{ad }e_{-1})\circ(\text{ad }e_0)\circ(\text{ad }e_{1})\nn\\
&=(\text{ad }f_0)\circ(\text{ad }f_1)\circ(\text{ad }f_0)\circ(\text{ad }f_{-1})
\nn\\&=
(\text{ad }f_0)\circ(\text{ad }f_{-1})\circ(\text{ad }f_0)\circ(\text{ad }f_{1}).
\end{align}
For $m=n$, we also factor out the one-dimensional ideal spanned by the Cartan element
\begin{align} \label{idealelement}
h_{-m}+2h_{-m+1}
+\cdots+mh_{-1}+(m+1)h_0+mh_{1}+\cdots+2h_{m-1}+h_m.
\end{align}

It follows from the Chevalley-Serre relations that $A(m,\,n)$ is spanned by all elements $e_{ij},\,f_{ij}$ and $h_{ij}$,
defined by
\begin{align} \label{amnbas}
e_{ij}&=[\cdots[e_i,\,e_{i+1}],\,\ldots,\,e_j],\nn\\
f_{ij}&=(-1)^j[\cdots[f_i,\,f_{i+1}],\,\ldots,\,f_j],\nn\\
h_{ij}&=h_{i}+h_{i+1}+\cdots+h_j,
\end{align}
where
$-m \leq i \leq j \leq n$. 
As for $C(n+1)$, the subspaces $\bfU_{-1}$ 
and $\bfU_1$ in the 3-grading $A(m,\,n)=\bfU_{-1}\oplus\bfU_0\oplus\bfU_1$ are spanned by all multiple commutators where $e_0$ and $f_0$, respectively, appear exactly once.
If $m\neq n$, then the Lie algebra $\bfU_0$ is the direct sum of a one-dimensional subalgebra and the complex semisimple Lie algebra $A_m \oplus A_n$. If $m=n$, the one-dimensional subalgebra vanishes, since this is then the ideal, spanned by the element (\ref{idealelement}), that we factored out in the construction of the algebra. The Dynkin diagram of $A_m \oplus A_n$ is obtained by deleting the grey node above, and its compact real form is $\su(m+1) \oplus \su(n+1)$.

In the basis (\ref{amnbas}), the subspaces $\bfU_1$ and $\bfU_{-1}$ are spanned by all elements $e_{ik}$ and $f_{jl}$, respectively, such that
$i,\,j\leq 0 \leq  k,\,l$.
Then we have
\begin{align} 
[h_{ik},\,e_{jl}]&=(-\de_{ij}+\de_{kl})e_{jl},\nn\\
[h_{ik},\,f_{jl}]&=(\de_{ij}-\de_{kl})f_{jl}. \label{forsteg}
\end{align}
If we furthermore assume $i < j$ and $k < l$, then we get
\begin{align}
\{e_{ik},\,f_{jk}\}&=-e_{i(j-1)},\nn\\
\{e_{jk},\,f_{ik}\}&=f_{i(j-1)},\nn\\
\{e_{jk},\,f_{jl}\}&=-f_{(k+1)l},\nn\\
\{e_{jl},\,f_{jk}\}&=-e_{(k+1)l}, \label{steg}
\end{align}
and we also have
\begin{align} \label{steg2}
\{e_{ij},\,f_{ij}\}&=h_{ij}.
\end{align}
For $(i,\,j)\neq(k,\,l)$, the anticommutator $\{e_{ij},\,f_{kl}\}$ is instead another element $e_{pq}$ or $f_{pq}$.
Using (\ref{forsteg})--(\ref{steg2}) we obtain
\begin{align} \label{formel}
[\{e_{ik},\,f_{jl}\},\,e_{pq}]&=
\de_{ij}\de_{lq}e_{kp}
-\de_{kl}\de_{jp}e_{iq}
\end{align}
for any $i,\,j\leq 0 \leq  k,\,l$.
Define the inner product $\kappa$ and the superconjugation $\tau$ by
\begin{align}
\kappa(h_i,\,h_j)&=-A_{ij},& \kappa(e_i,\,e_j)&=-\delta_{ij},
\end{align}
\begin{align}
\tau(e_q)&=f_q,& 
\tau(e_0)&=-f_0,&
\tau(e_p)&=-f_p,\nn\\
\tau(f_q)&=e_q,&
\tau(f_0)&=e_0,&
\tau(f_p)&=-e_p,\nn\\
\tau(h_q)&=-h_q,&
\tau(h_0)&=-h_0,&
\tau(h_p)&=-h_p,
\end{align}
for (only here) $q<0<p$.
This gives
\begin{align}
\kappa(e_{ik},\,f_{jl})&=-\delta_{ij}\delta_{kl}, &
\tau(e_{ik})&=-f_{ik},&\tau(f_{jl})&=e_{jl}
\end{align}
for $i,\,j\leq 0 \leq k,\,l$.
It then follows that the vector space $\bfU_{-1}$ together with the triple product given for the basis elements by
\begin{align} \label{trippelprodforbaselement}
(e_{ik}e_{jl}e_{pq})
=[\{ e_{ik},\,\tau(e_{jl})\},\,
e_{pq}]=
\delta_{kl}\de_{jp}e_{iq}-\de_{ij}\de_{lq}e_{kp}
\end{align}
is an $N=6$ three-algebra.

To each basis element $e_{ij}$ of $\bfU_{-1}$ we can associate an $m\times n$ matrix where the entry in row $i$ and column $j$ is one, and all the others are zero. Thus we can identify $\bfU_{-1}$ with the vector space of all complex $m\times n$ matrices. The triple product that follows from (\ref{trippelprodforbaselement})
for three arbitrary elements is then
\begin{align}
(xyz)=xy^\dagger z-zy^\dagger x.
\end{align}
We have thus shown that this example of an $N=6$ three-algebra, associated to the Lie superalgebra $A(m,\,n)$, is indeed the same as the one in \cite{Bagger:2008se}, where it was shown to give the ABJM theory. Alternatively, one could have shown this using the matrix realization of $A(m,\,n)$ as $\su(m+1|n+1,\,\mathbb{C})$ or (if $m=n$) as $\mathfrak{psu}(n+1|n+1,\,\mathbb{C})$, with (minus) the supertranspose as the superconjugation
$\tau$, and also obtained an analogous description of the three-algebra associated to 
$C(n+1)=\mathfrak{osp}(2|2n,\,\mathbb{C})$.

\section{Conclusion}\label{concl}

We have in this paper used generalized Jordan triple systems to describe a one-to-one correspondence between two different algebraic structures: on the one hand side the three-algebras that have been used in three-dimensional superconformal Chern-Simons theories, and on the other a kind of 3-graded Lie superalgebras.

As mentioned in the introduction, the correspondence between three-algebras and Lie superalgebras has also been studied in \cite{deMedeiros:2008zh}, without using generalized Jordan triple systems.
Theorem 22 in \cite{deMedeiros:2008zh} is similar to our Theorem 2.1, but there is an important difference. 
In \cite{deMedeiros:2008zh} the even subspace of the Lie superalgebra is required to act faithfully on the odd subspace in the adjoint representation, but it is not required to be simple. We have in this paper refined the result by showing that simple three-algebras correspond to simple Lie superalgebras and vice versa.

The simple Lie superalgebras that appear in the one-to-one correspondence are 
$C(n+1)$ and $A(m,\,n)$. We have only considered $N=6$ and $N=8$ theories, but there are also $N=4$ and $N=5$ theories corresponding to the Lie superalgebras 
$B(m,\,n)$, $D(m,\,n)$, $F(4)$, $G(3)$, $D(2,\,1;\,\alpha)$ \cite{Gaiotto:2008sd,Hosomichi:2008jd,Hosomichi:2008jb,Bergshoeff:2008bh} (for the embedding tensor approach see also \cite{Bergshoeff:2008ix}). These Lie superalgebras do not admit a 3-grading but instead a 5-grading, with nonzero subspaces $\bfU_{\pm2}$. It would be interesting to study the generalized Jordan triple systems corresponding to such 5-graded Lie superalgebras. Those corresponding to 5-graded ordinary Lie algebras are called \textit{Kantor triple systems} 
\cite{Kantor3.5,Allison,Mondocavhandling} and were suggested in \cite{Chowdhury:2009kk} to play a role in the Bagger-Lambert-Gustavsson theory of multiple M2-branes. 
In an other possibly related approach, {\em symplectic} three-algebras were introduced in \cite{Chen:2009ti}.
Going to the other side of the correspondence, the three-algebras that we have considered here are positive-definite, and this fact is essential for some of the results. Possible generalizations to other signatures would be an interesting subject for further research. 

Our construction of the associated Lie superalgebra is a simple modification of the original construction by Kantor, which associates a graded Lie algebra to any generalized Jordan triple system. These Lie algebras are defined by the same Cartan matrices (\ref{ccartan}) and (\ref{acartan}) as the corresponding Lie superalgebras, but the anticommutator in the Chevalley relations is replaced by a commutator. The result is an infinite-dimensional Lie algebra, which is not of Kac-Moody type \cite{Nilsson:2008kq}. However, the Lie algebra corresponding to $A(0,\,n)$ is in fact a \textit{Borcherds algebra}, which is a kind of generalized Kac-Moody algebra 
\cite{borcherds}. This would be interesting to explore further.
\newpage

\subsubsection*{Acknowledgments}
I would like to thank Bengt E.W.~Nilsson for inspiring discussions and encouragement. I am also grateful to
Jos\'e Figueroa-O'Farrill,
Carlo Meneghelli, Daniel Persson and Hidehiko Shimada for many valuable comments. Finally, I would like to thank the referee for making me aware of a few errors in the first version.


\appendix

\section{Simple three-algebras and Lie superalgebras}

As promised in section 2.3, we will here prove the remaining part of Theorem 2.1, that the 3-graded Lie superalgebra $\bfU$ associated to an $N=6$ three-algebra $V$ is simple if and only if $V$ is simple. First we need the following result, which we mentioned in section 2.2.

\begin{tjo}
Any weak ideal $D$ of an $N=6$ three-algebra $V$ is an ideal.
\end{tjo}
\Pf
Any two elements $u,\,v \in V$ can be decomposed as
\begin{align}
u &= u_D + u_\perp,& 
v &= v_D + v_\perp,
\end{align}
where $u_D,\,v_D \in D$, and 
\begin{align}
\la u_\perp | d \ra = \la v_\perp | d \ra = 0
\end{align}
for all $d \in D$.
Suppose that $D$ is not an ideal. Then, for any $d \in D$, there must be elements $u,\,v \in V$ such that $(udv) \notin D$. But since $D$ is a weak ideal,
\begin{align}
(u_D d v_D),\,(u_D d v_\perp),\,(u_\perp d v_D) \in D
\end{align}
so this means that $(u_\perp d v_\perp) \notin D$. Using (\ref{N=6ta3}) we get
\begin{align}
\la (u_\perp d v_\perp) | (u_\perp d v_\perp) \ra
=\la v_\perp | (du_\perp (u_\perp d v_\perp)) \ra = \la v_\perp | d' \ra 
\end{align}
for some $d' \in D$, again since $D$ is a weak ideal. Thus the inner product with $v_\perp$ is zero. But on the other hand, $(u_\perp d v_\perp)$ is a nonzero element and since the inner product is positive-definite, we must have
\begin{align}
\la (u_\perp d v_\perp) | (u_\perp d v_\perp) \ra > 0.
\end{align}
Thus we get a contradiction, and we conclude that $D$ is an ideal.
\qed

Since also the converse is true, any ideal is a weak ideal,
it follows from the lemma that it is enough to show that 
$\bfU$ is simple  if and only if 
there are no weak ideals of $V$ (other than 0 and $V$ itself). Furthermore, a weak ideal of $V$ is the same as a subrepresentation of $\bfU_{-1}$ under $\bfU_0$, so in fact we only have to show that
$\text{\it\textbf{U}}$ is simple if and only if 
the representation of $\text{\it\textbf{U}}_{0}$ on $\text{\it\textbf{U}}_{-1}$ is irreducible.

Suppose first that the representation of $\text{\it\textbf{U}}_{0}$ on $\text{\it\textbf{U}}_{-1}$ is irreducible, and that there is a nontrivial ideal $C$ of $\text{\it\textbf{U}}$. Let $c=c_{-1}+c_0+c_1$ be a nonzero element in $C$, where $c_k \in \text{\it\textbf{U}}_{k}$ for $k=0,\,\pm1$. 
It follows from the definition of $\bfU_0$ that it cannot contain a nonzero ideal of $\bfU$. Thus $c_{-1},\,c_1 \neq 0$
and $\tau(c_1) \in \text{\it\textbf{U}}_{-1}=V$. Since 
the inner product is non-degenerate, it follows from (\ref{N=6ta3}) that
there must be elements $u,\,v \in V$ such that $(u\tau(c_1)v)\neq0$. But
\begin{align} 
(u\tau(c_1)v) = [ \{ u,\,\tau(\tau(c_1)) \} ,\,v ] =
-[ \{ u,\,c_1 \} ,\,v ]
= -[\{ u,\,c\},\,v],
\end{align}
where the last equality follows from the 3-grading.
This is an element in $\text{\it\textbf{U}}_{-1}$, but also in $C$, since 
$C$ is an ideal of the Lie superalgebra. Thus the intersection of $\text{\it\textbf{U}}_{-1}$ and $C$ is a nonzero subrepresentation of $\text{\it\textbf{U}}_{-1}$
under $\text{\it\textbf{U}}_{0}$. But then $\text{\it\textbf{U}}_{-1} \subseteq C$ since the representation of $\text{\it\textbf{U}}_{0}$ on $\text{\it\textbf{U}}_{-1}$ is irreducible. In the same way, there must be elements $\tau(x),\,\tau(y) \in \bfU_1=\tau(V)$ such that $(\tau(x)\tau(c_{-1})\tau(y))\neq0$ and we get that $\text{\it\textbf{U}}_{-1} \subseteq C$. Since $\bfU$ is generated by
$\bfU_{-1}$ and $\bfU_1$, we conclude that $\bfU=C$, so the only nontrivial ideal of $\bfU$ is $\bfU$ itself. Thus $\bfU$ is simple.

Suppose now that $\bfU$ is simple, and let $D$ be a proper and nontrivial ideal of $V=\bfU_{-1}$, so that $[ \bfU_0,\, D] \subseteq D$. Then the subspace
\begin{align}
D + \{ D,\,\bfU_1 \} + [ \{ D,\,\bfU_1\},\,\bfU_1]
\end{align}
is a proper and nontrivial ideal of $\bfU$, which contradicts the assumption that $\bfU$ is simple. Thus there are no such ideals in $V$, 
and we conclude that $V$ is simple.




\begin{thebibliography}{10}

\bibitem{Schwarz:2004yj}
J.~H. Schwarz,  {\em {Superconformal Chern-Simons theories}}, JHEP {\bf 11},
  078 (2004)
[\href{http://www.arXiv.org/abs/hep-th/0411077}{{\tt hep-th/0411077}}].

\bibitem{Bagger:2006sk}
J.~Bagger and N.~Lambert,  {\em Modeling multiple {M}2's}, Phys. Rev. {\bf
  D75}, 045020 (2007)
[\href{http://www.arXiv.org/abs/hep-th/0611108}{{\tt hep-th/0611108}}].

\bibitem{Bagger:2007jr}
J.~Bagger and N.~Lambert,  {\em Gauge symmetry and supersymmetry of multiple
  {M}2-branes},
  Phys.\ Rev.\ {\bf D77} 065008 (2008) 
\href{http://www.arXiv.org/abs/{\rm[}arXiv:0711.0955 [hep-th]{\rm]}}{{\tt
  {\rm[}arXiv:0711.0955 [hep-th]{\rm]}}}.

\bibitem{Bagger:2007vi}
J.~Bagger and N.~Lambert,  {\em Comments on multiple {M}2-branes},
JHEP {\bf 02}, 105 (2008)
\href{http://www.arXiv.org/abs/{\rm[}arXiv:0712.3738~[hep-th]{\rm]}}{{\tt
  {\rm[}arXiv:0712.3738~[hep-th]{\rm]}}}.

\bibitem{Gustavsson:2007vu}
A.~Gustavsson,  {\em Algebraic structures on parallel {M}2-branes},
Nucl.\ Phys.\  {\bf B811}, 66 (2009)
\href{http://www.arXiv.org/abs/{\rm[}arXiv:0709.1260~[hep-th]{\rm]}}{{\tt
  {\rm[}arXiv:0709.1260~[hep-th]{\rm]}}}.

\bibitem{Papadopoulos:2008sk}
G.~Papadopoulos,  {\em {M2-branes, 3-Lie algebras and Pl\"ucker relations}},
  JHEP {\bf 05}, 054 (2008)
[\href{http://www.arXiv.org/abs/arXiv:0804.2662 [hep-th]}{{\tt arXiv:0804.2662
  [hep-th]}}].

\bibitem{Gauntlett:2008uf}
J.~P. Gauntlett and J.~B. Gutowski,  {\em {Constraining maximally
  supersymmetric membrane actions}},
  JHEP {\bf 06}, 053 (2008)
\href{http://www.arXiv.org/abs/{\rm[}arXiv:0804.3078~[hep-th]{\rm]}}{{\tt
  {\rm[}arXiv:0804.3078~[hep-th]{\rm]}}}.

\bibitem{Bagger:2008se}
J.~Bagger and N.~Lambert,  {\em {Three-algebras and N=6 Chern-Simons gauge
  theories}},
  Phys.\ Rev.\  {\bf D79}, 025002 (2009)
\href{http://www.arXiv.org/abs/{\rm[}arXiv:0807.0163~[hep-th]{\rm]}}{{\tt
  {\rm[}arXiv:0807.0163~[hep-th]{\rm]}}}.

\bibitem{Aharony:2008ug}
O.~Aharony, O.~Bergman, D.~L. Jafferis and J.~Maldacena,  {\em {N=6
  superconformal Chern-Simons-matter theories, M2-branes and their gravity
  duals}}, JHEP {\bf 10}, 091 (2008)
[\href{http://www.arXiv.org/abs/arXiv:0806.1218~[hep-th]}{{\tt
  arXiv:0806.1218~[hep-th]}}].

\bibitem{Benna:2008zy}
M.~Benna, I.~Klebanov, T.~Klose and M.~Smedb{\"a}ck,  {\em {Superconformal
  Chern-Simons theories and $AdS_4/CFT_3$ correspondence}}, JHEP {\bf 09}, 072
  (2008)
[\href{http://www.arXiv.org/abs/arXiv:0806.1519 [hep-th]}{{\tt arXiv:0806.1519
  [hep-th]}}].

\bibitem{Aharony:2008gk}
O.~Aharony, O.~Bergman and D.~L. Jafferis,  {\em {Fractional M2-branes}}, JHEP
  {\bf 11}, 043 (2008)
[\href{http://www.arXiv.org/abs/arXiv:0807.4924 [hep-th]}{{\tt arXiv:0807.4924
  [hep-th]}}].

\bibitem{Schnabl:2008wj}
M.~Schnabl and Y.~Tachikawa,  {\em {Classification of $N=6$ superconformal
  theories of ABJM type}},
\href{http://www.arXiv.org/abs/0807.1102}{[{\tt arXiv:0807.1102 [hep-th]}]}.

\bibitem{Gaiotto:2008sd}
D.~Gaiotto and E.~Witten,  {\em {Janus configurations, Chern-Simons couplings,
  and the theta-angle in N=4 super Yang-Mills theory}},
\href{http://www.arXiv.org/abs/{\rm[}arXiv:0804.2907~[hep-th]{\rm]}}{{\tt
  {\rm[}arXiv:0804.2907~[hep-th]{\rm]}}}.

\bibitem{Hosomichi:2008jd}
K.~Hosomichi, K.-M. Lee, S.~Lee, S.~Lee and J.~Park,  {\em {N=4 superconformal
  Chern-Simons theories with hyper and twisted hyper multiplets}}, JHEP {\bf
  07}, 091 (2008)
[\href{http://www.arXiv.org/abs/arXiv:0805.3662 [hep-th]}{{\tt arXiv:0805.3662
  [hep-th]}}].

\bibitem{Hosomichi:2008jb}
K.~Hosomichi, K.-M. Lee, S.~Lee, S.~Lee and J.~Park,  {\em {N=5,6
  superconformal Chern-Simons theories and M2-branes on orbifolds}}, JHEP {\bf
  09}, 002 (2008)
[\href{http://www.arXiv.org/abs/arXiv:0806.4977 [hep-th]}{{\tt arXiv:0806.4977
  [hep-th]}}].

\bibitem{Kantor3.5}
I.~L. Kantor,  {\em Some generalizations of {J}ordan algebras}, Trudy Sem.
  Vect. Tens. Anal. {\bf 16}, 407--499 (1972).

\bibitem{Nilsson:2008kq}
B.~E.~W. Nilsson and J.~Palmkvist,  {\em {Superconformal {M}2-branes and
  generalized {J}ordan triple systems}},
  Class.\ Quant.\ Grav.\  {\bf 26}, 075007 (2009)
\href{http://www.arXiv.org/abs/{\rm[}arXiv:0807.5134 [hep-th]{\rm]}}{{\tt
  {\rm[}arXiv:0807.5134 [hep-th]{\rm]}}}.

\bibitem{Bars:1978yx}
I.~Bars and M.~G{\"u}naydin,  {\em Construction of {L}ie algebras and {L}ie
  superalgebras from ternary algebras}, J. Math. Phys. {\bf 20}, 1977
(1979).

\bibitem{Kamiya1}
N.~Kamiya and S.~Okubo,  {\em Construction of {L}ie superalgebras
  ${D}(2,\,1;\,\alpha)$, ${G}(3)$ and ${F}(4)$ from some triple systems}, Proc.
  Edinburgh Math. Soc. {\bf 46}, 87--98 (2003).

\bibitem{Kamiya2}
N.~Kamiya and S.~Okubo,  {\em A construction of simple {L}ie superalgebras of
  certain types from triple systems}, Bull. Austral. Math. Soc. {\bf 69},
  113--123 (2004).

\bibitem{Kamiya3}
N.~Kamiya,  {\em A construction of {L}ie superalgebras ${B}(m,\,n)$ and
  ${D}(m,\,n)$ from triple systems}, Bulg. J. Phys. {\bf 33}, 128--139 (2006).

\bibitem{deMedeiros:2008zh}
P.~de~Medeiros, J.~Figueroa-O'Farrill, E.~Mendez-Escobar and P.~Ritter,  {\em
  {On the Lie-algebraic origin of metric 3-algebras}},
  Commun.\ Math.\ Phys.\  {\bf 290}, 871 (2009)
\href{http://www.arXiv.org/abs/{\rm[}arXiv:0809.1086 [hep-th]{\rm)}}{{\tt
  {\rm[}arXiv:0809.1086 [hep-th]{\rm]}}}.

\bibitem{Cherkis:2008qr}
S.~A. Cherkis and C.~S{\"a}mann,  {\em {Multiple M2-branes and generalized
  3-Lie algebras}}, Phys. Rev. {\bf D78}, 066019 (2008)
[\href{http://www.arXiv.org/abs/arXiv:0807.0808 [hep-th]}{{\tt arXiv:0807.0808
  [hep-th]}}].

\bibitem{Faulkner2}
J.~R. Faulkner,  {\em On the geometry of inner ideals}, J. Algebra {\bf 26},
  1--9 (1973).

\bibitem{Kac77A}
V.~G. Kac,  {\em A sketch of {L}ie superalgebra theory}, Comm. Math. Phys. {\bf
  53}, 31--64 (1977).

\bibitem{Kac77B}
V.~G. Kac,  {\em {L}ie superalgebras}, Adv. Math. {\bf 26}, 8--96 (1977).

\bibitem{FigueroaO'Farrill:2009pa}
  J.~Figueroa-O'Farrill,
  {\em Simplicity in the Faulkner construction},
  J.\ Phys.\ A  {\bf 42} (2009) 445206
  [{\tt arXiv:0905.4900 [hep-th]}].

\bibitem{deMedeiros:2009eq}
  P.~de Medeiros, J.~Figueroa-O'Farrill and E.~Mendez-Escobar,
{\em Superpotentials for superconformal Chern-Simons theories from representation theory},
  [{\tt arXiv:0908.2125 [hep-th]}].

\bibitem{Kantor-graded}
I.~L. Kantor,  {\em Graded {L}ie algebras}, Trudy Sem. Vect. Tens. Anal. {\bf
  15}, 227--266 (1970).

\bibitem{Frappat}
L.~Frappat, A.~Sciarrino and P.~Sorba, {\em Dictionary on Lie algebras and
  superalgebras}.
\newblock Academic Press, 2000.

\bibitem{Bergshoeff:2008bh}
E.~A. Bergshoeff, O.~Hohm, D.~Roest, H.~Samtleben and E.~Sezgin,\\{\em {The
  superconformal gaugings in three dimensions}}, JHEP {\bf 09}, 101 (2008)
[\href{http://www.arXiv.org/abs/arXiv:0807.2841 [hep-th]}{{\tt arXiv:0807.2841
  [hep-th]}}].

\bibitem{Bergshoeff:2008ix}
E.~A. Bergshoeff, M.~de~Roo, O.~Hohm and D.~Roest,  {\em {Multiple membranes
  from gauged supergravity}}, JHEP {\bf 08}, 091 (2008)
[\href{http://www.arXiv.org/abs/arXiv:0806.2584 [hep-th]}{{\tt arXiv:0806.2584
  [hep-th]}}].

\bibitem{Allison}
B.~N. Allison and J.~R. Faulkner,  {\em Elementary groups and invertibility for
  {K}antor pairs}, Comm. Algebra {\bf 27}, 519--556 (1999).

\bibitem{Mondocavhandling}
D.~Mondoc, {\em Kantor triple systems}.
\newblock Doctoral thesis, Lund University, 2002.

\bibitem{Chowdhury:2009kk}
S.~P. Chowdhury, S.~Mukhopadhyay and K.~Ray,  {\em {BLG theory with generalized
  Jordan triple systems}},
\href{http://www.arXiv.org/abs/{\rm[}arXiv:0903.2951 [hep-th]{\rm]}}{{\tt
  {\rm[}arXiv:0903.2951 [hep-th]{\rm]}}}.

\bibitem{Chen:2009ti}
  F.~M.~Chen and Y.~S.~Wu,
  {\em Symplectic three-algebra and ${\cal N}=6,\,Sp(2N)\times U(1)$ superconformal
  Chern-Simons-matter theory},
  [{\tt arXiv:0902.3454 [hep-th]}].


\bibitem{borcherds}
R.~E. Borcherds,  {\em Generalized {K}ac-{M}oody algebras}, J. Algebra {\bf
  115}, 501 (1988).

\end{thebibliography}

\providecommand{\href}[2]{#2}\begingroup\raggedright\endgroup

\end{document}